\documentclass[superscriptaddress,showpacs,12pt]{revtex4}
\usepackage{amsfonts}
\usepackage{graphicx, amsmath}
\usepackage{subfigure}
\begin{document}
\title[Short Title]{Counterfactual quantum-information transfer}
\author{Qi Guo}
\affiliation{Center for the Condensed-Matter Science and
Technology, Department of Physics, Harbin Institute of Technology,
Harbin, Heilongjiang 150001, People's Republic of China}
\author{Liu-Yong Cheng}
\affiliation{Center for the Condensed-Matter Science and
Technology, Department of Physics, Harbin Institute of Technology,
Harbin, Heilongjiang 150001, People's Republic of China}
\author{Li Chen}
\affiliation{Center for the Condensed-Matter Science and
Technology, Department of Physics, Harbin Institute of Technology,
Harbin, Heilongjiang 150001, People's Republic of China}
\author{Hong-Fu Wang}
\affiliation{Department of Physics, College of Science, Yanbian
University, Yanji, Jilin 133002, People's Republic of China}
\author{Shou Zhang\footnote{E-mail: szhang@ybu.edu.cn}}
\affiliation{Center for the Condensed-Matter Science and
Technology, Department of Physics, Harbin Institute of Technology,
Harbin, Heilongjiang 150001, People's Republic of China}
\affiliation{Department of Physics, College of Science, Yanbian
University, Yanji, Jilin 133002, People's Republic of China}
\begin{abstract}
We demonstrate quantum information can be transferred between two
distant participants without any physical particles travelling
between them. The key procedure of the counterfactual scheme is to
entangle two nonlocal qubits with each other without interaction,
so the scheme can also be used to generate nonlocal entanglement
counterfactually. We here illustrate the scheme by using flying
photon qubits and stationary electron-spin qubits assisted by
quantum dots inside double-sided optical microcavities. Unlike the
typical teleportation, the present scheme does not require prior
entanglement sharing or classical communication between the two
distant participants. 
\end{abstract}
\maketitle

\section{Introduction}

Quantum mechanics predicts many novel counterintuitive effects,
such as quantum entanglement, nonlocality, complementarity, and so
on. Combined with classical information science, quantum mechanics
promotes an interdisciplinary field in recent decades, i.e.
quantum information science \cite{1}, which can achieve lots of
information processing tasks that appear unimaginable in the
classical domain. In quantum information, the minimal unit is
qubit, which is usually encoded in the quantum state of a physical
entity. Hence the transfer of quantum state carrying quantum
information, i.e. quantum information transfer, is the foundation
of quantum communication. In 1993, Bennett \emph{et al.} proposed
that an unknown quantum state can be teleported to a distant
receiver with the help of prior entanglement sharing and classical
communication \cite{2}. That scheme, called quantum teleportation,
has opened the door for the intense study of quantum communication
\cite{3,4,5}. Another strategy for transferring an unknown quantum
state to a distant location can be achieved by using a flying
qubit to interact with two spatially separated stationary qubits
\cite{6,7,8}. Note that these kinds of quantum information
transfer scheme require the particles carrying information
(classical bits or qubits) to travel between the separated
participants.

On the other hand, counterfactual quantum information processing
has been attracting more and more scientists' attention in recent
years. The counterfactuality means relevant quantum information
tasks can be achieved without physical particles travelling
between two parties. The research in this aspect originated from
the interaction-free measurements proposed by Elitzur and Vaidman
in 1993 \cite{9}, whose basic idea was that an obstructing object
in one of the arms of the Mach-Zehnder interferometer could
destroy the interference even if no photon was absorbed by the
object. Thus, one can ascertain the existence of the object in the
given arm of the interferometer with the maximum attainable
efficiency 50\%, though no photon ``touched" this object. In 1995,
Kwiat \emph{et al.} improved the interaction-free measurements
\cite{10} and the probability of an interaction-free measurement
could be made arbitrarily close to 100\% by applying a discrete
form of the quantum Zeno effect \cite{11}, which refers to one
coherently repeats the interrogation of the region that might
contain the object. Using a novel ``chained" version of the
quantum Zeno effect, Hosten \emph{et al.} demonstrated
counterfactual quantum computation and implemented Grover's search
algorithm \cite{12} with boosting the counterfactual inference
probability to unity. In 2009, Noh proposed a counterfactual
quantum key distribution (CQKD) scheme based on quantum
interrogation \cite{13}, where the secret information could be
distributed in a secure way between two remote parties even though
no particle transmitted through the quantum channel. Subsequently,
the unconditional security of the CQKD was proved in an ideal
situation \cite{14}, and this CQKD scheme was realized
experimentally \cite{15}. Very recently, Salih \emph{et al.}
presented a direct counterfactual quantum communication protocol
\cite{16}, in the ideal asymptotic limit, which allowed a
classical bit to be transferred from the sender to the receiver
without any particles travelling between them by using the
``chained" quantum Zeno effect. And this work has also attracted
much attention \cite{17,18}.

We note that, in all counterfactual schemes above, the presence or
absence of the obstructing object in one of the arms of the
interferometer was classically controlled by the experimenter. In
this paper, we examine what happens if we replace the classical
control with a quantum control device. The results show that when
let the obstructing object be in an unknown quantum superposition
state of presence and absence, it can be counterfactually
transferred to a distant place, which extremely differs from the
typical teleportation protocol. It may be more intuitive to
explain the principle of our scheme by using the chained
Mach-Zehnder type interferometer similarly to Ref.~\cite{16}.
However, considering the effectiveness of resources and the
experimental feasibility, we will elaborate the scheme with nested
Michelson-type interferometer. In principle, as long as the
quantum control device can be implemented, our scheme is universal
for many physical systems of quantum information processing such
as trapped ion systems, superconducting quantum systems, and so
on. In order to illustrate the scheme in detail, here we take the
quantum dot (QD) double-sided optical microcavity system for
example, in which the interaction between flying photon qubits and
stationary electron-spin qubits has been well studied in
solid-state quantum computation and communication\cite{19,20,21}.

The paper is organized as follows. In
Sec.~\uppercase\expandafter{\romannumeral2}, we briefly introduce
the construction of a spin-cavity unit and construct a quantum
control device using the unit. In
Sec.~\uppercase\expandafter{\romannumeral3}, we illustrate how to
implement the transfer of quantum information between two distant
participants without any physical particles travelling between
them. In Sec.~\uppercase\expandafter{\romannumeral4}, we
numerically analyze and discuss the effect of imperfections of the
experimental conditions for the present scheme. A conclusion is
given in Sec.~\uppercase\expandafter{\romannumeral5}.

\section{Quantum control device implemented by a singly charged QD in a double-sided microcavity}

Now, we discuss how to control the blocking or passing of photons
by a quantum state, which is equivalent to placing the obstructing
object of the previous counterfactual schemes in a quantum
superposition state of presence and absence. We implement the
quantum control device by QD-microcavity system. Consider a system
consisting of a singly charged self-assembled GaAs/InAs QD being
embedded in an optical resonant double-sided microcavity, which
proposed by Hu \emph{et al.} \cite{19} and recognized by Bonato
\emph{et al.} \cite{20} recently.  The charged exciton $X^{-}$,
produced by the optical excitation of the system, consists of two
electrons bound in one hole. The four relevant electronic levels
are shown in Fig.~1 (a), where the symbols $\Uparrow$
$(\Downarrow)$ and $\uparrow$ $(\downarrow)$ represent a heavy
hole and an electron with $z$-direction spin projections
$+\frac{3}{2}$ $(-\frac{3}{2})$ and $+\frac{1}{2}$
$(-\frac{1}{2})$, respectively. The two electrons in the exciton
are in a singlet state, which means the two electrons have total
spin zero, thus, the electron-spin interactions with the heavy
hole spin are avoided. The excess electron spin in the QD
interacts with the cavity mode through the addition of a charged
exciton. According to the optical selection rules and the
transmission and reflection rules of the cavity for an incident
circular polarization photon, the interactions between photons and
electrons in the QD-microcavity coupled system can be described as
follows \cite{20}:
\begin{eqnarray}\label{e1}
&&|R^{\uparrow}, \uparrow\rangle\rightarrow|L^{\downarrow},
\uparrow\rangle, ~~~|L^{\uparrow},
\uparrow\rangle\rightarrow-|L^{\uparrow}, \uparrow\rangle,
\cr&&|R^{\downarrow}, \uparrow\rangle\rightarrow-|R^{\downarrow},
\uparrow\rangle, ~~~|L^{\downarrow},
\uparrow\rangle\rightarrow|R^{\uparrow}, \uparrow\rangle,
\cr&&|R^{\uparrow}, \downarrow\rangle\rightarrow-|R^{\uparrow},
\downarrow\rangle, ~~~|L^{\uparrow},
\downarrow\rangle\rightarrow|R^{\downarrow}, \downarrow\rangle,
\cr&&|R^{\downarrow}, \downarrow\rangle\rightarrow|L^{\uparrow},
\downarrow\rangle, ~~~|L^{\downarrow},
\downarrow\rangle\rightarrow-|L^{\downarrow}, \downarrow\rangle,
\end{eqnarray}
where $|R\rangle$ and $|L\rangle$ denote the right-circularly
polarized photon state and the left-circularly polarized photon
state, respectively, and the superscript uparrow (downarrow)
denotes the propagating direction of polarized photon along
(against) the $z$ axis.

\begin{figure}
\renewcommand\figurename{\small Fig.}
 \centering \vspace*{8pt} \setlength{\baselineskip}{10pt}
 \subfigure[]{
 \includegraphics[scale = 1.2]{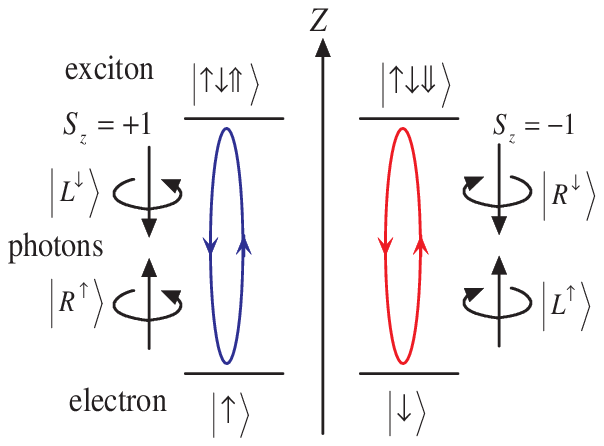}}
 \subfigure[]{
 \includegraphics[scale = 1.2]{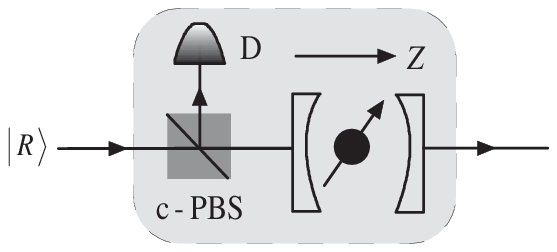}}
 \caption{\label{f1} (a) Relevant energy levels and optical selection
 rules for the optical transition of negatively charged exciton in
 GaAs/InAs quantum dots embedded in an optical microcavity.
 The superscript arrows of the photon states indicates their propagation
 direction along or against the $z$ axis. (b) Quantum control device for the passing
 or blocking of the incident right-circularly polarized photons. c-PBS denotes the polarizing beam splitter which transmits
the right circularly polarized photon $|R\rangle$ and reflects the
left circularly polarized photon $|L\rangle$. D: conventional
photon detector.}
\end{figure}

From Eq.~(1), we can see that, for an incident photon with spin
$s_{z}=+1$ ($|R^{\uparrow}\rangle$ or $|L^{\downarrow}\rangle$),
if the electron is in the state $|\uparrow\rangle$, the photon
will couple with the electron and be reflected by the cavity. Then
the photon state is transformed into the state
$|L^{\downarrow}\rangle$ or $|R^{\uparrow}\rangle$, respectively,
that is, both the photon's polarization and the propagation
direction are flipped. On the other hand, if the electron is in
the state $|\downarrow\rangle$, there is no dipole interaction and
the photon is transmitted through the cavity and acquires a $\pi$
mod $2\pi$ phase shift relative to a reflected photon. Similarly,
a photon with spin $s_{z}=-1$ ($|R^{\downarrow}\rangle$ or
$|L^{\uparrow}\rangle$) will be reflected when the electron-spin
state is $|\downarrow\rangle$ and will be transmitted through the
cavity when the electron-spin state is $|\uparrow\rangle$.

Using the above transmission and reflection rules of the
photon-QD-microcavity system, we can construct a quantized
obstructing object, i.e. the passing or blocking of the incident
photon is controlled by an unknown quantum state rather than the
experimenter. Take the right-circularly polarized photon state
$|R\rangle$ for example, the quantum device is shown in Fig.~1(b).
The electron in a QD is initially in an arbitrary superposition
state
$|\varphi\rangle=\alpha|\uparrow\rangle+\beta|\downarrow\rangle$.
Let a right-circularly polarized photon first passe through a
polarizing beam splitter in the circular basis (c-PBS), which
transmits the right circularly polarized photon and reflects the
left circularly polarized photon. So the photon will enter into
the cavity along the $z$ axis. According to Eq.~(1), if the
electron is in the state $|\uparrow\rangle$, the photon will
couple with the QD, then be reflected by the cavity and become
left-circularly polarized photon simultaneously. So after
reflected by c-PBS, the photon will be absorbed by the detector D.
While, if the electron is in the state $|\downarrow\rangle$, the
photon will pass through the cavity. In short, the
right-circularly polarized photon will be absorbed for electron
state $|\uparrow\rangle$, and it will pass through the cavity for
electron state $|\downarrow\rangle$. Therefore, the shaded area in
Fig.~1(b) works as the quantum version of the obstructing object
only for right-circularly polarized photons, and the blocking or
passing of the photon depends on the electron's quantum state,
which is equivalent to that the obstructing object is in the
superposition of presence and absence.

\section{The scheme of counterfactual unknown quantum state transfer}

We will show the counterfactual quantum state transfer with a
nested Michelson-type interferometer and QD-microcavity unit in
this section. The nested Michelson-type interferometer is
constructed by embedding a Michelson-type interferometer into one
of the arms of another interferometer.

\subsection{Partially counterfactual nonlocal entangled state
generation}

\begin{figure}
\scalebox{1.5}{\includegraphics{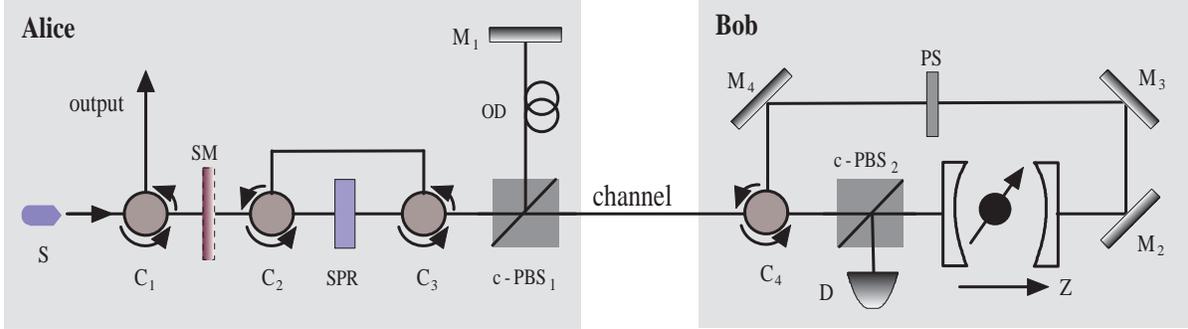}}\caption{\label{f2}
Schematic of partially counterfactual nonlocal entangled state
generation. \emph{S} stands for the light source, $C_{i}$
($i=1,2,3,4$) is the optical circulator. SM: switchable mirror.
SPR: switchable polarization rotator, which rotates the
polarization by an angle $\theta$. $\mathrm{M}_{i}$ ($i=1,2,3,4$)
is normal mirror. c-PBS: polarizing beam splitter in the circular
basis. D: conventional photon detector. PS is the phase shifter
used to perform the transformation
$|R\rangle\leftrightarrow-|R\rangle$. }
\end{figure}

Before discussing the counterfactual quantum information transfer,
it is necessary to introduce a method to generate nonlocal
polarization-spin entangled state by repeatedly using a
Michelson-type interferometer with a QD-microcavity inserted in
one of the arms, which is used as the inner interferometer in the
following quantum state transfer scheme. The setup is shown in
Fig.~2, where SM indicates the switchable mirror that can be
switched on and off by external means, and M is normal mirror. The
two optical paths SM$\rightarrow\mathrm{M}_{1}$ and
SM$\rightarrow\mathrm{M}_{3}$ form a Michelson-type
interferometer. Optical delay (OD) is used to match the optical
path lengthes of the different paths of the interferometer. The
light source in Alice's site emits left circularly polarized
photons $|L\rangle$, and the excess electron in Bob's QD is in an
arbitrary state $\alpha|\uparrow\rangle+\beta|\downarrow\rangle$.
Initially, the SM is switched off, i.e. allowing the photon to be
transmitted. Once the photon enters in the interferometer, the SM
is switched on (reflects the photon) and remains on during the
photon travels $N$ cycles in the interferometer. So after the
photon enters into the interferometer, it is firstly rotated an
angle $\theta$ by the switchable polarization rotator (SPR), whose
action is given by the transformation
$|L\rangle\rightarrow\cos\theta|L\rangle+\sin\theta|R\rangle$ and
$|R\rangle\rightarrow\cos\theta|R\rangle-\sin\theta|L\rangle$. The
state of the polarization-spin hybrid system becomes
\begin{eqnarray}\label{e2}
|\varphi\rangle_{0}\rightarrow(\cos\theta|L\rangle+\sin\theta|R\rangle)(\alpha|\uparrow\rangle+\beta|\downarrow\rangle).
\end{eqnarray}
Then the photon passes through a c-PBS. Thus, the $|L\rangle$
component of the photon will be propagated to the normal mirror
$\mathrm{M}_{1}$ and stays in Alice's site, and the $|R\rangle$
component of the photon will be propagated to Bob's site and
injected into the optical microcavity to interact with the QD
spin. We can see that only the right circularly polarized photon
can be injected into the microcavity along the $z$ axis of the
QD-microcavity, i.e. only the interactions \{$|R^{\uparrow},
\uparrow\rangle\rightarrow|L^{\downarrow}, \uparrow\rangle$,
$|R^{\uparrow}, \downarrow\rangle\rightarrow-|R^{\uparrow},
\downarrow\rangle$\} in Eq.~(1) may occur in the present scheme.
So in the following the superscript arrows of the photon state are
omitted. After the interaction, the state of the system evolves as
\begin{eqnarray}\label{e3}
|\varphi\rangle_{0}\rightarrow\cos\theta|L\rangle(\alpha|\uparrow\rangle+\beta|\downarrow\rangle)+\alpha\sin\theta|L\rangle|\uparrow\rangle-\beta\sin\theta|R\rangle|\downarrow\rangle.
\end{eqnarray}
It can be seen the $|L\rangle$ component produced by interaction
in Bob's site will be reflected by c-PBS$_{2}$ and absorbed by the
detector D, and the $|R\rangle$ component will come back to the SM
in Alice's site by optical elements. The phase shifter (PS) in
Bob's site provides a $\pi$ phase shift for the right circularly
polarized photon state, $|R\rangle\rightarrow-|R\rangle$.
Therefore, when the detector does not click and the photon comes
back to the SM, i.e. after the first cycle of the photon
travelling in the interferometer, the state is given by
\begin{eqnarray}\label{e4}
|\varphi\rangle_{1}&=&\cos\theta|L\rangle(\alpha|\uparrow\rangle+\beta|\downarrow\rangle)+\beta\sin\theta|R\rangle|\downarrow\rangle
\cr&=&\alpha\cos\theta|L\rangle|\uparrow\rangle+(\cos\theta|L\rangle+\sin\theta|R\rangle)\beta|\downarrow\rangle.
\end{eqnarray}
Note that the above state is not normalized, because the
$|L\rangle$ component in Bob's site is absorbed by the detector
and the component $\alpha\sin\theta|L\rangle|\uparrow\rangle$ in
Eq~(3) is ignored here. In the same way, after $N$ cycles, the SM
is switched off to allow the photon to be transmitted and exit
from the output port, and the system state becomes
\begin{eqnarray}\label{e5}
|\varphi\rangle_{N}=\alpha\cos^{N}\theta|L\rangle|\uparrow\rangle+(\cos
N\theta|L\rangle+\sin N\theta|R\rangle)\beta|\downarrow\rangle.
\end{eqnarray}
Let $\theta=\pi/2N$, the final state is
\begin{eqnarray}\label{e6}
|\varphi\rangle_{N}=\alpha\cos^{N}\frac{\pi}{2N}|L\rangle|\uparrow\rangle+\beta|R\rangle|\downarrow\rangle,
\end{eqnarray}
which is a non-maximal polarization-spin hybrid entangled state.
The probability of obtaining the entangled state is
$|\alpha|^{2}\cos^{2N}(\pi/2N)+|\beta|^{2}$. Obviously, for the
large cycles $N$, the probability will be close to unit and the
state will be normalized,
$|\varphi\rangle_{N}\sim\alpha|L\rangle|\uparrow\rangle+\beta|R\rangle|\downarrow\rangle$,
and for $\alpha=\beta=1/\sqrt{2}$, the nonlocal maximal entangled
state can be obtained. The present protocol may not be the optimal
scheme for the nonlocal entangled state preparation, however, it
is worth noting that the present scheme is partially
counterfactual. During the process of the entangled-state
generation, although the photon passes through the transmission
channel between Alice and Bob, the photon doesn't interact with
Bob's electron at all. From the discussions above, once the photon
interacts with the electron, it will be reflected by the cavity
and be absorbed by the detector. Therefore, this is an
interaction-free scheme for nonlocal entangled state generation.
In the following, we will show completely counterfactual entangled
state generation and quantum state transfer without any particles
travelling in the transmission channel.

\subsection{Completely counterfactual nonlocal entangled
state generation and quantum state transfer}

\begin{figure}
\scalebox{1.2}{\includegraphics{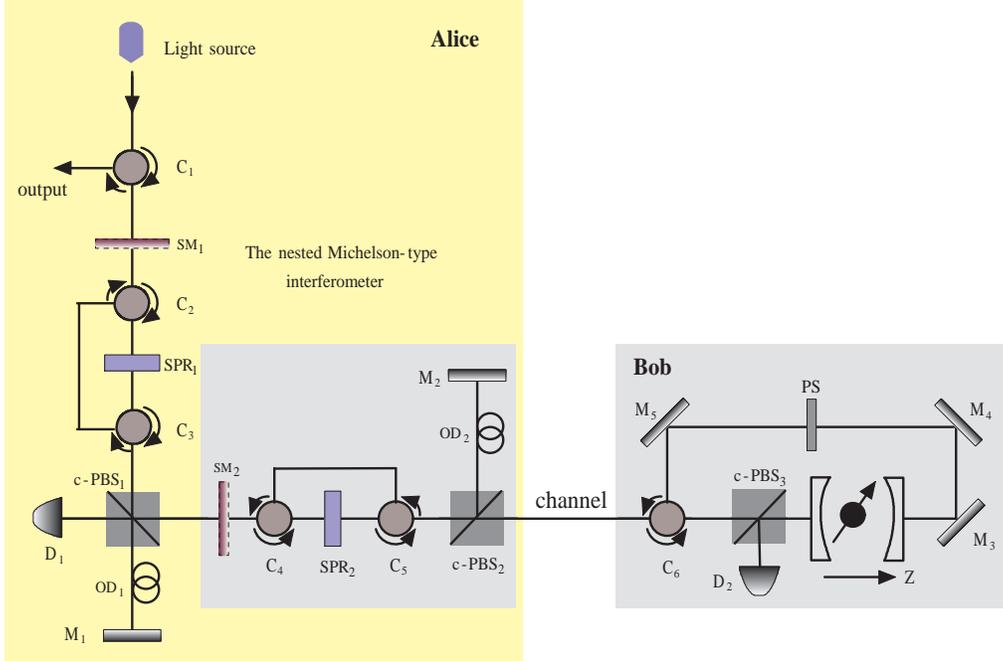}}\caption{\label{f3} The
nested Michelson interferometer used to implement counterfactual
nonlocal entangled state generation and quantum state transfer.
All the optical elements are the same as ones in Fig.~2. The setup
in the gray area is the interferometer in Fig.~2, which acts as an
inner interferometer inserted in one of the arms of an outer
Michelson interferometer. }
\end{figure}

Now we show how to counterfactually generate a nonlocal entangled
state and transfer an unknown quantum state from Bob to Alice
without any physical particles travelling between them. The scheme
is accomplished in the ideal limit, by using the nested Michelson
interferometer, which is shown in Fig.~3. The interferometer in
Fig.~2, as an inner interferometer, is inserted in one of the arms
of an outer Michelson interferometer. The two optical paths
$\mathrm{SM}_{1}\rightarrow \mathrm{M}_{1}$ and
$\mathrm{SM}_{1}\rightarrow \mathrm{M}_{3}$ form the outer
interferometer, which means the photon must undergo $N$ inner
cycles in every outer cycle. Choosing suitable cycle numbers $N$
and $M$ corresponding to the inner interferometer and outer
interferometer respectively, an unknown quantum state transfer can
be counterfactually achieved with probability close to unit.
Suppose sender Bob wants to transfer an arbitrary quantum state
$\alpha|\uparrow\rangle+\beta|\downarrow\rangle$ of the electron
in QD to receiver Alice. Alice sends a right circularly polarized
photon $|R\rangle$ into the interferometer for the input port. The
photon passes through the $\mathrm{SM}_{1}$, which is switched off
initially (transmits photons), and after the photon enters into
the interferometer, the $\mathrm{SM}_{1}$ remains on (reflects the
photon) until the photon finishes the $M$th cycle in the outer
interferometer. The $\mathrm{SPR}_{1}$ performs the
transformations
$|L\rangle\rightarrow\cos\vartheta|L\rangle+\sin\vartheta|R\rangle$
and
$|R\rangle\rightarrow\cos\vartheta|R\rangle-\sin\vartheta|L\rangle$
($\vartheta=\pi/2M$), and the joint state of the photon and the
electron becomes
\begin{eqnarray}\label{e7}
|\psi\rangle_{0}&\rightarrow&(\cos\vartheta|R\rangle-\sin\vartheta|L\rangle)(\alpha|\uparrow\rangle+\beta|\downarrow\rangle)
\cr&=&\cos\vartheta|R\rangle(\alpha|\uparrow\rangle+\beta|\downarrow\rangle)-\sin\vartheta|L\rangle(\alpha|\uparrow\rangle+\beta|\downarrow\rangle).
\end{eqnarray}
Because of the c-PBS$_{1}$, the $|R\rangle$ component of the state
will be transmitted, while the $|L\rangle$ component will be
reflected and injected into the inner interferometer for going
through the $N$ inner cycles. The evolution of the second term of
Eq.~(7) in the inner interferometer is the same as the above
subsection. After the $N$ cycles in the inner interferometer, the
system state is given by
\begin{eqnarray}\label{e8}
|\psi\rangle_{0}\rightarrow\cos\vartheta|R\rangle(\alpha|\uparrow\rangle+\beta|\downarrow\rangle)-\sin\vartheta(\alpha\cos^{N}\frac{\pi}{2N}|L\rangle|\uparrow\rangle+\beta|R\rangle|\downarrow\rangle).
\end{eqnarray}
When the photon comes back from the inner interferometer and
$\mathrm{M}_{1}$, the $|R\rangle$ component coming from the inner
interferometer will be absorbed by the detector $\mathrm{D}_{1}$.
So at the end of the first cycle in the outer interferometer, the
state can be written as
\begin{eqnarray}\label{e9}
|\psi\rangle_{1}=\cos\vartheta|R\rangle(\alpha|\uparrow\rangle+\beta|\downarrow\rangle)-\alpha\sin\vartheta\cos^{N}\frac{\pi}{2N}|L\rangle|\uparrow\rangle.
\end{eqnarray}
Both Eq.~(8) and Eq.~(9) are not normalized, because the
components absorbed by the detectors $\mathrm{D}_{1}$ and
$\mathrm{D}_{2}$ can't reach the SMs and be ignored. Through
calculation we know when the photon finishes the $m$th ($2\leq
m\leq M$) outer cycle, the system state can be written as
\begin{eqnarray}\label{e10}
|\psi\rangle_{m}=\alpha x_{m}|R\rangle|\uparrow\rangle+\beta
y_{m}|R\rangle|\downarrow\rangle-\alpha
z_{m}|L\rangle|\uparrow\rangle,
\end{eqnarray}
where the parameters $x_{m}$, $y_{m}$, and $z_{m}$ satisfy the
recursion relations
\begin{eqnarray}\label{e11}
&&x_{m}=x_{m-1}\cos\vartheta-z_{m-1}\sin\vartheta, \cr
&&y_{m}=y_{m-1}\cos\vartheta, \cr
&&z_{m}=(x_{m-1}\sin\vartheta+z_{m-1}\cos\vartheta)\cos^{N}\frac{\pi}{2N},
\end{eqnarray}
with $x_{1}=y_{1}=\cos\vartheta$ and
$z_{1}=\sin\vartheta\cos^{N}(\pi/2N)$. When the photon finishes
$M$ outer cycles, the SM$_{1}$ is switched off and the photon
exits from the output port. We plot the variation trend of the
parameters $x_{m}$, $y_{m}$, and $z_{m}$ with the values of $N$
and $M$, as shown in Fig.~4. It's obvious that
$x_{m}\rightarrow0$, $y_{m}\rightarrow1$, and $z_{m}\rightarrow1$
for large values of $N$ and $M$, for example, ($x=0.0059$,
$y=0.9994$, and $z=0.9905$) for ($M=30$ and $N=2000$), which means
that after the $M$th outer cycle, the nonlocal hybrid entangled
state can be obtained with the probability close to 1. That is
\begin{eqnarray}\label{e12}
|\psi\rangle_{M}\simeq\beta|R\rangle|\downarrow\rangle-\alpha|L\rangle|\uparrow\rangle.
\end{eqnarray}
So far, the nonlocal entangled state generation is achieved.
Obviously, during the whole process, the probability that the
photon travels the channel is nearly suppressed to 0 by repeatedly
using the nested Michelson interferometer. In other words, as long
as the photon passes through the channel, it will be absorbed by
either $\mathrm{D}_{1}$ or $\mathrm{D}_{2}$ on the way back from
Bob to Alice. So this is a counterfactual scheme with no photon
passing through the transmission channel.

\begin{figure}
\renewcommand\figurename{\small Fig.}
 \centering \vspace*{8pt} \setlength{\baselineskip}{10pt}
 \subfigure[]{
 \includegraphics[scale = 0.35]{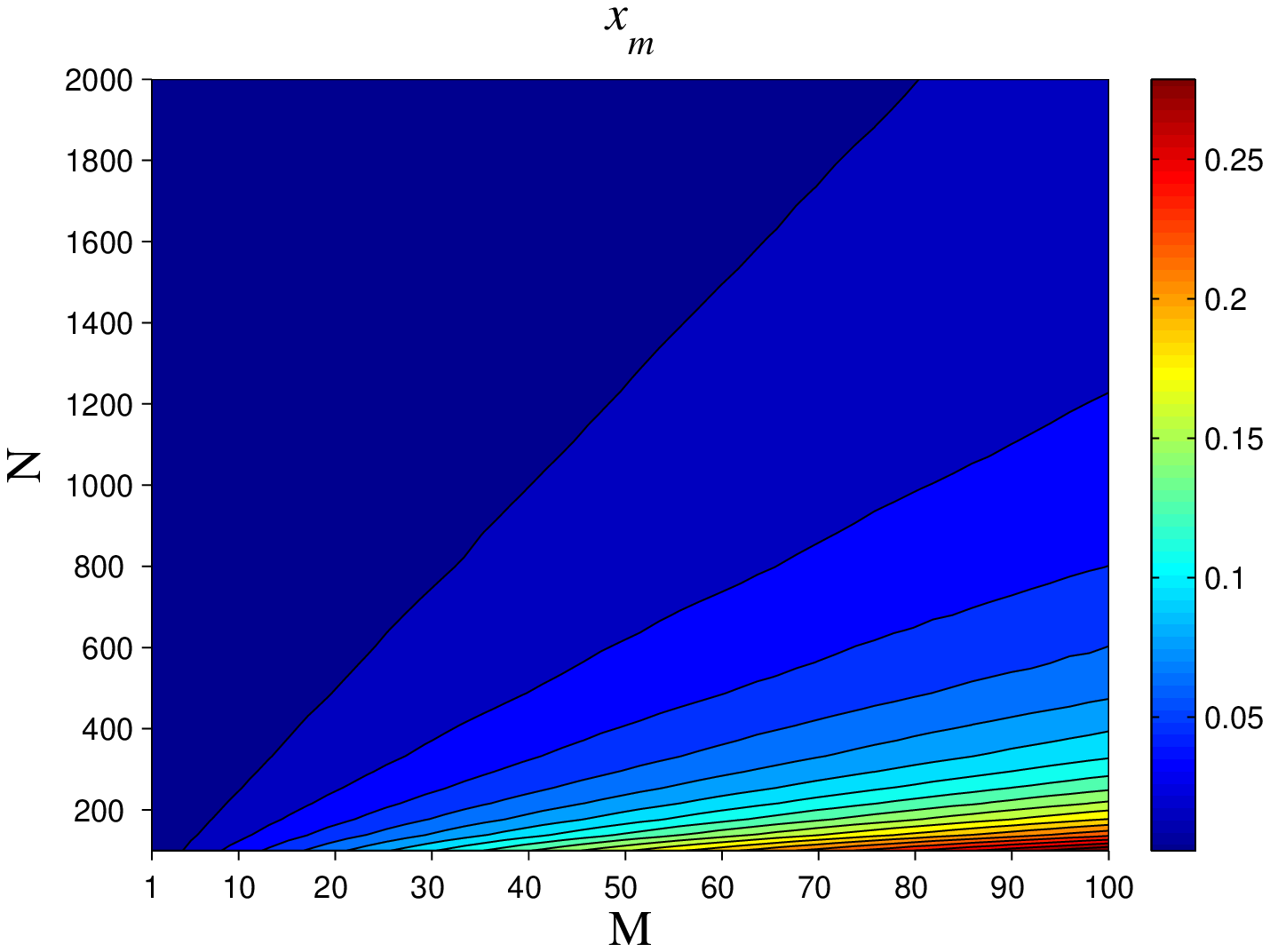}}
 \subfigure[]{
 \includegraphics[scale = 0.35]{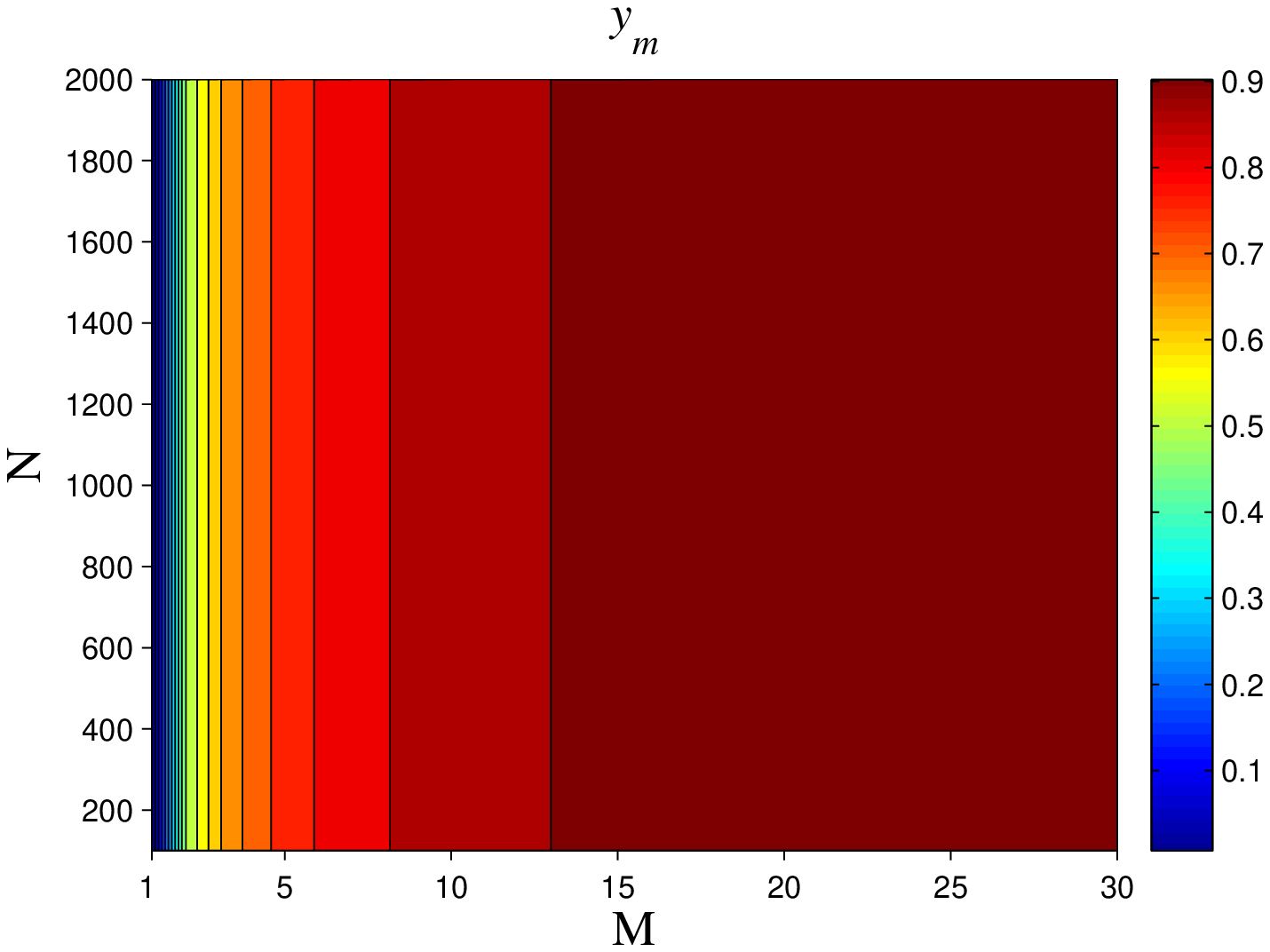}}
  \subfigure[]{
 \includegraphics[scale = 0.35]{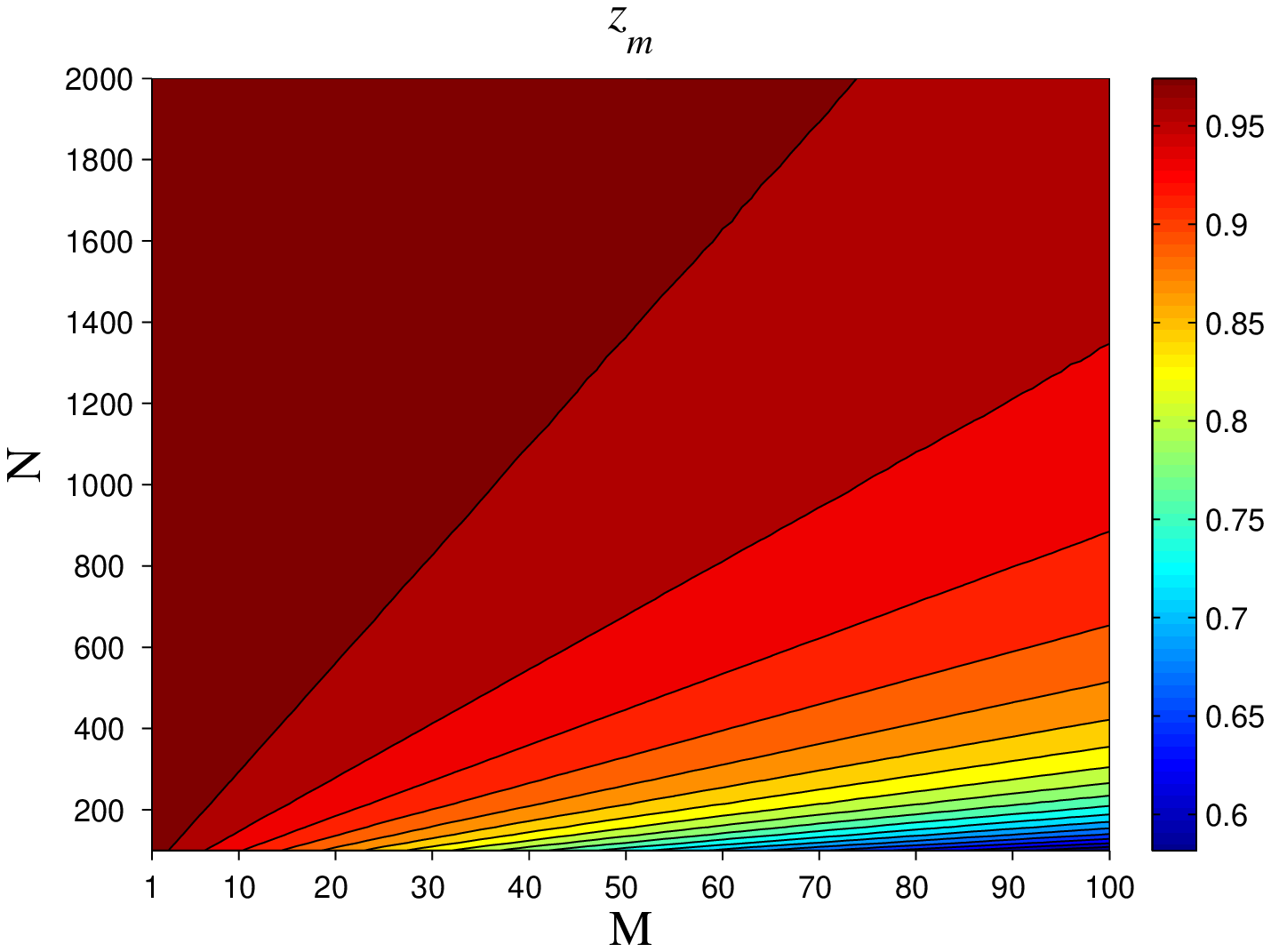}}
 \caption{\label{f3} The parameters $x_{m}$, $y_{m}$, and $z_{m}$ in Eq.~(10) versus the different values of $N$
and $M$. (a) $x_{m}$ is close to zero for large $N$ and
appropriate $M$. (b) $y_{m}$ approaches 1 with the increase of $M$
and doesn't change with $N$. (c) $z_{m}$ is close to 1 for
appropriate values of $N$ and $M$.}
\end{figure}

It's straightway to achieve the quantum state transfer, once the
entangled state in Eq.~(12) is obtained. Bob performs a Hadamard
transformation
\{$|\uparrow\rangle\rightarrow(|\uparrow\rangle+|\downarrow\rangle)/\sqrt{2}$,

$|\downarrow\rangle\rightarrow(|\uparrow\rangle-|\downarrow\rangle)/\sqrt{2}$\}
on the electron state, which can be implemented by using a $\pi/2$
microwave or optical pulse \cite{22,23,24}, then the state is
given by
\begin{eqnarray}\label{e13}
|\psi\rangle_{M}\rightarrow-\frac{1}{\sqrt{2}}[(\alpha|L\rangle-\beta|R\rangle)|\uparrow\rangle-(\alpha|L\rangle+\beta|R\rangle)|\downarrow\rangle].
\end{eqnarray}
Then Bob detects the spin in the basis \{$|\uparrow\rangle$,
$|\downarrow\rangle$\}. If the detection result isn't informed
Alice by classical communication, Alice can obtain the teleported
state with the probability of 50\%, however, if the detection
result is sent to Alice, she will perfectly obtain the transferred
state with the help of a single-qubit phase flip gate.

It has been shown that a quantum state (or qubit) can be
transferred without exchanging particle between the two
participants. Compared with the typical quantum teleportation, the
present scheme doesn't require prior entanglement sharing or even
classical communication. Even if Alice wants to obtain the quantum
information deterministically, Bob only needs to send one bit
information to Alice for the transfer of one qubit, on the
contrary, two bits classical communication is required in the
one-qubit teleportation procedure \cite{2}. Moreover, while the
quantum state is transferred to the photon, the initial state of
the electron is destroyed by Bob's detection, which makes the
scheme avoid to violate the quantum no-cloning theorem. On the
other hand, during the state transfer process, although the photon
does not travel to Bob's site, the optical path length it travels
is near $2MN$ times that of the distance between Alice and Bob, so
the scheme here cannot realize the superluminal communication.
Therefore, the present scheme achieves the quantum
counterfactuality without contradicting any existing physical law.

\section{Analysis and discussion}

Now we analyze and discuss the performance of the quantum
information transfer. Obviously, this scheme can be accomplished
with the probability close to 100\% under the ideal conditions.
However, considering the practical experimental implementation of
the present scheme, the performance must be affected by the
imperfections of the system. First, the scheme requires
high-precision switchable polarization rotators $\mathrm{SPR}_{1}$
and $\mathrm{SPR}_{2}$ to rotate a single photon state by angles
$\vartheta=\pi/(2M)$ and $\theta=\pi/(2N)$ respectively, which
however are bound to be introduced a slight error in the practical
situations. As defined in Ref.~\cite{16}, we suppose the error
coefficient of the $\mathrm{SPR}_{1(2)}$ is $s_{1(2)}$, which
indicates the photon state is rotated with an additional angle
$\Delta\vartheta=s_1\vartheta/M$ ($\Delta\theta=s_2\theta/N$)
after each outer (inner) cycle. Therefore, we can derive the real
final state $|\psi\rangle^{'}_{M}$ after the $M$th outer cycle by
replacing the rotated angles $\vartheta$ ($\theta$) in the
recursion relations of Eq.~(11) with $\vartheta+\Delta\vartheta$
($\theta+\Delta\theta$). In order to estimate the influence of the
SPRs error, we analyze the average fidelity of the system state
after $M$ outer cycles. Without loss of generality, let the
normalization coefficients $\alpha$ and $\beta$ in Eq.~(12) equal
$\cos\xi$ and $\sin\xi$, respectively. And the average fidelity of
the final state can be written as
$\mathcal{\overline{F}}=\frac{1}{2\pi}\int_{0}^{2\pi}d\xi|_{M}\langle\psi|\psi\rangle_{M}^{'}|^{2}$.
Assume the error coefficients $s_1=s_2=s$, we numerically estimate
the average fidelity and plot its change with $s$ for different
values of $M$ and $N$ in Fig.~5, which indicates that the fidelity
is higher for lesser error factor $s$.
\begin{figure}
\scalebox{0.6}{\includegraphics{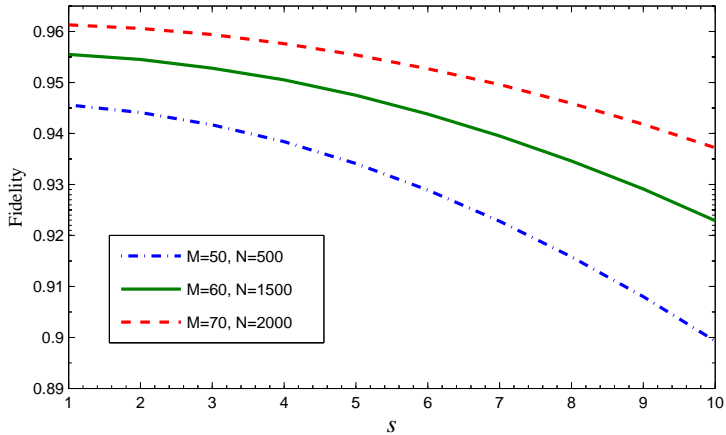}}\caption{\label{f5} The
average fidelity of the counterfactual quantum state transfer
versus the error coefficient $s$ of the switchable polarization
rotators. }
\end{figure}

Other crucial influence factors come from the spin-cavity system
that is the key component in the present scheme. It seems that the
performance of the spin-cavity unit does not affect the
counterfactual quantum state transfer scheme since no photon
passes through transmission channel to interact with the quantum
dot during the whole process. However, the spin-cavity system can
affect the scheme through influencing the efficiency of the inner
interferometer. Therefore, we can estimate the effect coming from
the performance of the spin-cavity system by analyzing the
successful probability of the partially counterfactual entangled
state generation in
Sec.~\uppercase\expandafter{\romannumeral3}(A). The reflection and
transmission coefficients of coupled and the uncoupled cavities
are generally different when the side leakage and the cavity loss
are not negligible, which has been illustrated in the
Ref.~\cite{21}. Under the assumption of weak excitation limit, the
reflection and transmission coefficients of a double-sided optical
microcavity are described by \cite{19,21}
\begin{eqnarray}\label{e14}
&&r(\omega)=\frac{[i(\omega_{X^{-}}-\omega)+\frac{\gamma}{2}][i(\omega_{c}-\omega)+\frac{\kappa_{s}}{2}]+g^{2}}{[i(\omega_{X^{-}}-\omega)+\frac{\gamma}{2}][i(\omega_{c}-\omega)+\kappa+\frac{\kappa_{s}}{2}]+g^{2}},
\cr\cr&&t(\omega)=\frac{-\kappa[i(\omega_{X^{-}}-\omega)+\frac{\gamma}{2}]}{[i(\omega_{X^{-}}-\omega)+\frac{\gamma}{2}][i(\omega_{c}-\omega)+\kappa+\frac{\kappa_{s}}{2}]+g^{2}},
\end{eqnarray}
where $g$ is the coupling strength between $X^{-}$ and the cavity
field; $\omega$, $\omega_{c}$, and $\omega_{X^{-}}$ are
respectively the frequencies of the input photon, the cavity
field, and the $X^{-}$ transition; $\kappa$, $\kappa_{s}$, and
$\gamma$ are the cavity field decay rate, the side leakage rate,
and the $X^{-}$ dipole decay rate, respectively. For the case that
the QD does not interact with the input photon, i.e. $g=0$, under
the resonant interaction condition
$\omega_{c}=\omega_{X^{-}}=\omega_{0}$, the reflection and
transmission coefficients become
\begin{eqnarray}\label{e15}
&&r_{0}(\omega)=\frac{[i(\omega_{0}-\omega)+\frac{\kappa_{s}}{2}]}{[i(\omega_{0}-\omega)+\kappa+\frac{\kappa_{s}}{2}]},
\cr\cr&&t_{0}(\omega)=\frac{-\kappa}{[[i(\omega_{c}-\omega)+\kappa+\frac{\kappa_{s}}{2}]}.
\end{eqnarray}
Therefore, considering the side leakage and the cavity loss, the
rules of optical transitions in a realistic $X^{-}$-cavity system
used in this paper become \cite{19,20,21}
\begin{eqnarray}\label{e16}
&&|R^{\uparrow},\uparrow\rangle\rightarrow|r(\omega)||L^{\downarrow},\uparrow\rangle+|t(\omega)||R^{\uparrow},\uparrow\rangle,
\cr&&|R^{\uparrow},\downarrow\rangle\rightarrow-|t_{0}(\omega)||R^{\uparrow},\downarrow\rangle-|r_{0}(\omega)||L^{\downarrow},\downarrow\rangle.
\end{eqnarray}
Using above transition rules, we can calculate the probability of
generating the entangled state in Eq.~(6) under the presence of
side leakage and cavity loss. In order to illuminate the effect of
the spin-cavity system, we plot the probability of entangled state
generation versus the side leakage rate $\kappa_{s}/\kappa$ and
the normalized coupling strength $g/\kappa$ by means of numerical
simulation in the case of $\omega_{0}=\omega$ and $N=300$, as
shown in Fig.~6, which indicates that the present scheme is
especially sensitive to the side leakage rate but the effect of
the coupling strength is negligible. The numerical result is not
difficult to understand by reviewing the partially counterfactual
entangled state generation process, in which the photon passes
through $N$ cycles (here $N=300$) in the inner interferometer but
does not interact with the electron in the QD. It is clear that
the scheme requires optical cavity with extremely high quality
factor, for example, the probability only equals 0.63 for
$\kappa_{s}=0.01\kappa$ and $g=3\kappa$, which is a strong
challenge for side leakage in experiments \cite{25}. Fortunately,
the improvement of fabrication techniques can suppress the side
leakage \cite{26}. When the side leakage can be neglected compared
with the main cavity decay, i.e. $\kappa_{s}\ll\kappa$, the
probability will approach unity.

\begin{figure}
\scalebox{0.7}{\includegraphics{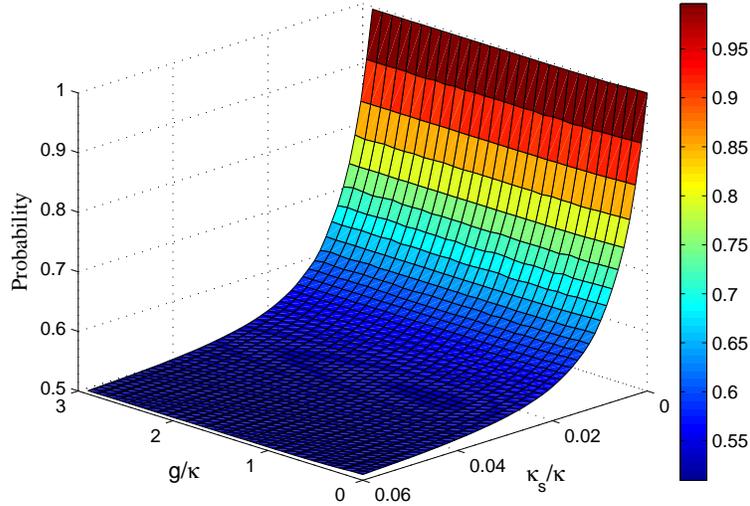}}\caption{\label{f6} The
successful probability of partially counterfactual nonlocal
entangled state generation using the inner interferometer versus
the side leakage rate $\kappa_{s}/\kappa$ and the normalized
coupling strength $g/\kappa$. Here we have set $\gamma=0.1\kappa$,
which is experimentally achievable. }
\end{figure}

In addition, the electron spin qubit used in the present scheme
and its fast initialization have been demonstrated \cite{27}, and
the photon qubit can be produced with the well-known spontaneous
parametric down-conversion \cite{28,29}. However, the background
noise will be produced during the process of photons' generation,
transmission, and detection, which is larger than the single
photon signal. In order to accomplish the scheme successfully, the
experiment should be performed under the condition with high
signal-noise ratio. Some recent works \cite{30,31} have
demonstrated the unwanted background photons can be efficiently
reduced in the parametric down-conversion process by using the
optical shutter controlled by a simple field programable gate
array. The background noise can also be suppressed by optimizing
the detuning between the frequencies of the pump and photon pairs
and cooling the nonlinear fiber \cite{31}. We also note that the
two detectors in our scheme are only used to absorb the photon
passed through the transmission channel, therefore, if the scheme
is completed successfully, it means the detectors detected no
photon in the whole process. In this sense, the sensitivity and
dark counts of the detectors do not influence the efficiency of
the scheme, and the detectors can even be replaced with other
absorption objects.

From the discussion and analysis above, the present scheme is
implementable in principle, but there may be some challenges for
current experiment conditions in practice. Nevertheless, compared
with the experimental difficulties and the practical applications
of the present scheme, perhaps the mind-boggling and highly
counter-intuitive fact of transferring quantum information without
transferring any particles is more worthy to pay attention, and
the deeper physical mechanism behind the counterfactual protocol
may be more interesting. Very recently, Vaidman \emph{et al.}
\cite{32} reported a series of interesting works about the past of
photons passing through a nested Mach-Zehnder interferometer.
Based on the locality of physical interactions, they proposed a
method to analyze the past of a photon via the weak trace it
leaves, which can be explained in the two-state vector formulation
(TSVF) of quantum mechanics \cite{33}. The results showed that the
past of photons can't be represented by continuous trajectories
and should be described by both forward- and backward-evolving
quantum states. That means a photon may be in every place between
its emission and detection points. These works maybe provide a
line of thinking for explaining the counterfactuality of the
present scheme, i.e. considering photons as delocalized waves with
discontinuous trajectories in the interferometer, i.e. considering
photons as delocalized waves with discontinuous trajectories in
the interferometer. On the other hand, perhaps the present
counterfactual scheme can also provide evidence for the viewpoint
in Ref.~\cite{32}. This issue is interesting, and is worth
studying further in the future.

\section{Conclusions}

In conclusion, we have proposed a counterfactual scheme for
transferring an unknown quantum state. The scheme indicated that a
qubit can be teleported to a distant place without any physical
particles travelling between them. The present scheme does not
require prior entanglement sharing and classical communication
between the two distant participants, so it essentially differs
from the typical teleportation. In addition, it is necessary to
entangle two nonlocal qubits without interaction during the
process of the quantum state transfer, thus the scheme can also be
used to generate nonlocal entanglement counterfactually. We also
numerically estimated the effect of the imperfections of the
experiment system, which indicated our scheme required optical
microcavity with high quality factor.

\begin{center}$\mathbf{Acknowledgments}$\end{center}

This work is supported by the National Natural Science Foundation
of China under Grant Nos. 61068001 and 11264042; China
Postdoctoral Science Foundation under Grant No. 2012M520612; the
Program for Chun Miao Excellent Talents of Jilin Provincial
Department of Education under Grant No. 201316; and the Talent
Program of Yanbian University of China under Grant No. 950010001.


\begin{thebibliography}{999}
\bibitem{1} M. A. Nielsen and I. L. Chuang, \emph{Quantum computation and quantum information} (Cambridge university press, Cambridge, U.K., 2000).
\bibitem{2} C. H. Bennett, G. Brassard, C. Cr\'{e}peau, R. Jozsa, A. Peres, and W. K. Wootters, Phys. Rev. Lett. \textbf{70}, 1895 (1993).
\bibitem{3} D. Bouwmeester, J. W. Pan, K. Mattle, M. Eibl, H. Weinfurter, and A. Zeilinger, Nature (London) \textbf{390}, 575 (1997).
\bibitem{4} A. Furusawa, J. L. S\o rensen, S. L. Braunstein, C. A. Fuchs, H. J. Kimble, and E. S. Polzik, Science \textbf{282}, 706 (1998).
\bibitem{5} M. A. Nielsen, E. Knill, and R. Laflamme, Nature (London) \textbf{396}, 52 (1998).
\bibitem{6} J. I.Cirac, P. Zoller, H. J. Kimble, and H. Mabuchi, Phys. Rev. Lett. \textbf{78}, 3221 (1997).
\bibitem{7} D. N. Matsukevich and A. Kuzmich, Science \textbf{306} 663 (2004).
\bibitem{8} M. A. Sillanp\"{a}\"{a}, J. I. Park, and R. W. Simmonds, Nature (London) \textbf{449}, 438 (2007).
\bibitem{9} A. C. Elitzur and L. Vaidman, Found. Phys. \textbf{23}, 987 (1993).
\bibitem{10} P. Kwiat, H. Weinfurter, T. Herzog, A. Zeilinger, and M. A. Kasevich, Phys. Rev. Lett. \textbf{74}, 4763 (1995).
\bibitem{11} B. Misra and E. C. G. Sudarshan, J. Math. Phys. \textbf{18}, 756 (1977).
\bibitem{12} O. Hosten, M. T. Rakher, J. T. Barreiro, N. A. Peters, and P. G. Kwiat, Nature (London) \textbf{439}, 949 (2006).
\bibitem{13} T.-G. Noh, Phys. Rev. Lett. \textbf{103}, 230501 (2009).
\bibitem{14} Z. Q. Yin, H. W. Li, W. Chen, Z. F. Han, and G. C. Guo, Phy. Rev. A \textbf{82}, 042335 (2010); Z. Q. Yin, H. W. Li, Y. Yao, C. M. Zhang, S. Wang, W. Chen, G. C. Guo, and Z. F. Han, \emph{ibid}. \textbf{86}, 022313 (2012).
\bibitem{15} Y. Liu, L. Ju, X. L. Liang, S. B. Tang, G. L. Shen Tu, L. Zhou, C. Z. Peng, K. Chen, T. Y. Chen, Z. B. Chen, and J. W. Pan, Phys. Rev. Lett. \textbf{109}, 030501 (2012); G. Brida, A. Cavanna, I.P. Degiovanni, M. Genovese, and P.Traina, Laser Phys. Lett. \textbf{9}, 247 (2012).
\bibitem{16} H. Salih, Z. H. Li, M. Al-Amri, and M. S. Zubairy, Phys. Rev. Lett. \textbf{110}, 170502 (2013).
\bibitem{17} N. Gisin, Phys. Rev. A \textbf{88}, 030301 (2013).
\bibitem{18} J. L. Zhang, F. Z. Guo, F. Gao, B. Liu, and Q. Y. Wen, Phys. Rev. A \textbf{88}, 022334 (2013).
\bibitem{19} C. Y. Hu, W. J. Munro, J. L. O'Brien, and J. G. Rarity, Phys. Rev. B \textbf{80}, 205326 (2009).
\bibitem{20} C. Bonato, F. Haupt, S. S. R. Oemrawsingh, J. Gudat, D. Ding, M. P. van Exter, and D. Bouwmeester, Phys. Rev. Lett. \textbf{104}, 160503 (2010).
\bibitem{21} C. Y. Hu and J. G. Rarity, Phys. Rev. B \textbf{83}, 115303 (2011).
\bibitem{22} D. Press, T. D. Ladd, B. Y. Zhang, and Y. Yamamoto, Nature (London) \textbf{456}, 218 (2008).
\bibitem{23} J. Berezovsky, M. H. Mikkelsen, N. G. Stoltz, L. A. Coldren, and D. D. Awschalom, Science \textbf{320}, 349 (2008).
\bibitem{24} A. Greilich, S. E. Economou, S. Spatzek, D. R. Yakovlev, D. Reuter, A. D. Wieck, T. L. Reinecke and M. Bayer, Nature Phys. \textbf{5}, 262 (2009).
\bibitem{25} J. P. Reithmaier, G. Sek, A. L¡§offler, C. Hofmann, S. Kuhn, S. Reitzenstein, L. V. Keldysh, V. D. Kulakovskii, T. L. Reinecke, and A. Forchel, Nature (London) \textbf{432}, 197(2004).
\bibitem{26} S. Reitzenstein, C. Hofmann, A. Gorbunov, M. Strau\ss, S. H. Kwon, C. Schneider, A. L\"{o}ffler, S. H\"{o}fling, M. Kamp, and A. Forchel, Appl. Phys. Lett. \textbf{90}, 251109 (2007).
\bibitem{27} C. Emary, X. Xu, D. G. Steel, S. Saikin, and L. J. Sham, Phys. Rev. Lett. \textbf{98}, 047401 (2007); D. Kim, S. E. Economou, S. C. Badescu, M. Scheibner, A. S. Bracker,M. Bashkansky, T. L. Reinecke, and D. Gammon, \emph{ibid}. \textbf{101}, 236804 (2008); D. Press, T. D. Ladd, B. Zhang, and Y. Yamamoto, Nature (London) 456, 218 (2008).
\bibitem{28} J.-W. Pan, Z.-B. Chen, C.-Y. Lu, H. Weinfurter, A. Zeilinger, and M. \.{Z}ukowski, Rev. Mod. Phys. \textbf{84}, 777 (2012).
\bibitem{29} P. G. Kwiat, K. Mattle, H. Weinfurter, A. Zeilinger, A. V. Sergienko, and Y. Shih,  Phys. Rev. Lett. \textbf{75}, 4337 (1995).
\bibitem{30} G. Brida, I. P. Degiovanni, M. Genovese, A. Migdall, F. Piacentini, S. V. Polyakov, and I. R. Berchera, Opt. Express \textbf{19}, 1484 (2011).
\bibitem{31} L. Yang, F. Sun, N. Zhao, and X. Li, Opt. Express \textbf{22}, 2553 (2014).
\bibitem{32} L. Vaidman, Phys. Rev. A \textbf{87}, 052104 (2013); A. Danan, D. Farfurnik, S. Bar-Ad, and L. Vaidman, Phys. Rev. Lett. \textbf{111}, 240402 (2013); L. Vaidman, Phys. Rev. A \textbf{89}, 024102 (2014).
\bibitem{33} Y. Aharonov, P. G. Bergmann, and J. L. Lebowitz, Phys. Rev. \textbf{134}, B1410 (1964); Y. Aharonov and L. Vaidman, Phys. Rev. A \textbf{41}, 11 (1990).
\end{thebibliography}
\end{document}